\documentclass{Interspeech2024}
\usepackage{graphicx}
\usepackage{amsmath,amssymb}
\usepackage{nicefrac}
\usepackage{tikz}
\usepackage{multirow}
\usepackage{makecell}
\usepackage[shortcuts]{extdash}
\usepackage{xspace}
\usepackage{placeins}
\usepackage{todonotes}
\usepackage[super]{nth}

\usepackage{pifont}
\newcommand{\cmark}{\ding{51}}%
\newcommand{\xmark}{\ding{55}}%

\newcommand{\fb}[0]{FluencyBank\xspace}
\newcommand{\sep}[0]{SEP\=/28k\xspace}
\newcommand{\sepE}[0]{SEP\=/28k\=/E\xspace}
\newcommand{\ksof}[0]{KSoF\xspace}

\interspeechcameraready

\title{Large Language Models for Dysfluency Detection in Stuttered Speech}

\name[affiliation={1}]{Dominik}{Wagner}
\name[affiliation={1}]{Sebastian P.}{Bayerl}
\name[affiliation={1}]{Ilja}{Baumann}
\name[affiliation={1}]{Korbinian}{Riedhammer}
\name[affiliation={2}]{Elmar}{Nöth}
\name[affiliation={1,3}]{Tobias}{Bocklet}

\address{
  $^1$Technische Hochschule Nürnberg Georg Simon Ohm\\
  $^2$Friedrich-Alexander-Universität Erlangen-Nürnberg \\
  $^3$Intel Labs}
\email{dominik.wagner@th-nuernberg.de}


\definecolor{nodefill}{RGB}{218,232,252}
\definecolor{nodedraw}{RGB}{108,142,191}
\newcommand*\circled[1]{\tikz[baseline=(char.base)]{
            \node[fill=nodefill,shape=circle,draw=nodedraw,inner sep=0.3pt] (char) {#1};}
            }

\let\OLDthebibliography\thebibliography
\renewcommand\thebibliography[1]{
  \OLDthebibliography{#1}
  \setlength{\parskip}{0pt}
  \setlength{\itemsep}{0pt plus 0.3ex}
}

\begin{document}

\maketitle

\begin{abstract}
Accurately detecting dysfluencies in spoken language can help to improve the performance of automatic speech and language processing components and support the development of more inclusive speech and language technologies. 
Inspired by the recent trend towards the deployment of large language models (LLMs) as universal learners and processors of non-lexical inputs, such as audio and video, we approach the task of multi-label dysfluency detection as a language modeling problem. 
We present hypotheses candidates generated with an automatic speech recognition system and acoustic representations extracted from an audio encoder model to an LLM, and finetune the system to predict dysfluency labels on three datasets containing English and German stuttered speech. 
The experimental results show that our system effectively combines acoustic and lexical information and achieves competitive results on the multi-label stuttering detection task. 
\end{abstract}
\vspace{-2mm}
\section{Introduction}
\vspace{-1mm}
Stuttering is a diverse neurodevelopmental condition that affects an individual's verbal communication ability \cite{bloodstein_handbook_2021} and can negatively impact the performance of automatic speech and language processing systems \cite{mitra21_interspeech,shaomai2023stutter,colin2023stutter}. 
The symptoms of stuttering vary strongly among individuals and depend on psychological influences, conversational factors, and the linguistic complexity of an utterance \cite{bloodstein_handbook_2021}. 
The accurate detection of dysfluent speech has implications for stuttering therapy, e.g., in self-therapy software, monitoring of stuttering by clinicians, as well as the development of more inclusive speech technologies in general. 
Dysfluency detection has traditionally been focused on acoustic features, since the various dysfluency types are often directly visible in audio waveforms and their corresponding spectrograms \cite{sheikh2022stuttering_review,barrett2022stutter}. 
Most studies employ spectral features like MFCCs or coefficients derived from linear predictive coding methods \cite{lea2021sep28k,jouaiti2022stutter}. 
Recent work has explored neural representations obtained from acoustic encoder models \cite{montacie2022stutter,bayerl22b_interspeech,bayerl2022partitioning,bayerl23_interspeech}, in particular wav2vec~2.0 \cite{baevski2020w2v2}. 
These representations are then used as input features to various classifiers ranging from support vector machines to neural networks. 

Lexical features such as transcriptions generated by automatic speech recognition (ASR) systems are not extensively studied in stuttering-related dysfluency detection, particularly in combination with acoustic features. 
However, they are commonly used to detect and remove dysfluencies in transcriptions obtained from typical speech \cite{zayats16_interspeech,qian2020disfluency}. 
Other studies use different feature types depending on the dysfluency class. 
For example, in \cite{alharbi18_interspeech}, prolongations are detected based on the correlation between successive audio frames and other stuttering-related dysfluencies are detected using word lattices.

While most studies on dysfluency detection focus on using
either lexical or acoustic features, it is well known that combinations of the two can improve the detection of speaking style differences in spoken dialog systems \cite{shriberg12_interspeech} and in distinguishing machine-directed speech from background speech \cite{wagner2023multimodal}. 

Large language models (LLMs) such as  Llama~2 \cite{touvron2023llama2}, Falcon \cite{almazrouei2023falcon} and GPT-4 \cite{openai2023gpt4} exhibit strong language understanding and generation abilities. 
More importantly, new capabilities have emerged that are not present in smaller-scale language models (e.g., multi-step reasoning, in-context learning, and instruction following) \cite{minaee2024large}. 
Furthermore, the potential of LLMs is not limited to text data alone. 
Recent studies have demonstrated their capability to process and understand non-lexical information such as audio, video, and images \cite{deshmukh2023pengi,tang2024salmonn,gong2024listen,chen2024its,rubenstein2023audiopalm}, bridging the gap between different modalities.

This work is inspired by the recent successful attempts to enable learning from non-lexical features in LLMs and their ability to adapt to new tasks. 
We explore a system to detect stuttering-related dysfluencies using three datasets of English and German stuttered speech. 
Our goal is to correctly identify all dysfluency labels in the data (i.e., blocks, interjections, prolongations, sound repetitions, and word repetitions), as well as modified speech in short time intervals. 
Besides the lexical inputs obtained through ASR transcriptions (either orthographic \cite{radford2022whisper} or phonetic \cite{xu2021w2v2phoneme}), we employ latent features generated with an acoustic encoder based on wav2vec 2.0. 
The ASR hypotheses candidates and acoustic features are concatenated and jointly presented to a pretrained  Llama 2 model, enabling it to consider both acoustic and lexical content. 
The system is optimized using low-rank-adaption (LoRA) \cite{hu2022lora} to generate dysfluency labels based on the acoustic and lexical context. 
Lexical features represent the characteristics of dysfluent speech at resolutions distinct from those of acoustic features. 
Acoustic features are fine-grained and able to capture occurrences of dysfluencies in narrow time frames, whereas lexical features are coarse and focus on self-contained entities (e.g. words). 
Therefore, we hypothesize that lexical and acoustic features are complementary, each offering distinct advantages in enhancing dysfluency detection. 
\vspace{-5mm}
\section{Data}
\vspace{-1mm}
For our experiments, we employ three datasets consisting of audio clips lasting 3 seconds each, annotated with stuttering-related dysfluencies; SEP-28k-Extended, \fb, and the Kassel State of Fluency (\ksof) dataset \cite{bayerl2022partitioning,lea2021sep28k,bayerl_KSoFKasselState_2022a}.  
The datasets have similar labels with clips containing either no dysfluencies or one or more types of stuttering-related dysfluencies; blocks, prolongations, sound repetitions, word repetitions, and interjections.
The clips can be labeled with more than one type of dysfluency, making it a multi-label classification problem.

The largest dataset, \sepE, comprises approximately $\sim$28k English audio clips sourced from podcasts discussing stuttering. 
Derived from the \sep dataset, it features semi-automatically generated speaker labels and a speaker-exclusive Train-Dev-Test partition.\footnote{Online: \protect\url{https://tinyurl.com/yck9fmfv}}
The original \sep release contains a subset of the adults who stutter dataset of the \fb corpus \cite{bernsteinratner_FluencyBankNew_2018} segmented into 4144 English clips that were annotated to match the annotations used in the larger dataset.
In our experiments, we employ the partition outlined in \cite{bayerl22b_interspeech}.\footnote{\protect Online: \url{https://tinyurl.com/24vm6dec}}

The \ksof dataset consists of 5597 audio segments extracted from German stuttering therapy recordings. 
In addition to the five dysfluency labels in \sepE and \fb, the clips can also contain annotated speech modifications, indicating that a person uses fluency shaping.
Fluency shaping is a technique persons who stutter learn in stuttering therapy to help them overcome their stuttering \cite{bayerl_KSoFKasselState_2022a}.
A detailed description of the datasets and the distribution of the labels can be found in \cite{lea2021sep28k,bayerl_KSoFKasselState_2022a,bayerl2022partitioning}.
\vspace{-5mm}
\section{Method}
\vspace{-2mm}
\subsection{wav2vec~2.0}
\vspace{-1mm}
Wav2vec~2.0 \cite{baevski2020w2v2} describes a series of models consisting of a convolutional feature encoder $\mathcal{G}: \boldsymbol{X}_{1:T} \mapsto \boldsymbol{Z}_{1:L}$ with multiple identical blocks using temporal convolution, layer normalization, and GELU nonlinearity. 
The feature encoder maps a raw audio waveform $\boldsymbol{X}_{1:T} = \lbrace x_1, \ldots , x_T \rbrace$ of length $T$ to hidden representations $\boldsymbol{Z}_{1:L} = \lbrace \boldsymbol{z}_1, \ldots , \boldsymbol{z}_L \rbrace$ of length $L$. 
These hidden representations $\boldsymbol{Z}$ are then passed to a transformer \cite{vaswani17Transformer} $\mathcal{T}: \boldsymbol{Z}_{1:L} \mapsto \boldsymbol{C}_{1:L}$, which generates context representations $\boldsymbol{c}_1, \ldots , \boldsymbol{c}_L$. 
During pretraining, a quantization component is used to quantize $\boldsymbol{Z}_{1:L}$, which serve as the targets in a contrastive learning task that requires classifying the true quantized representation within a set of distractors.  
Latent representations extracted at different transformer layers of wav2vec~2.0 are known to be reliable acoustic features for dysfluency detection \cite{bayerl22b_interspeech,bayerl23_interspeech} and other related tasks such as mispronunciation detection \cite{xu21k_interspeech}.
Pretrained wav2vec~2.0 models are available in two sizes (94M and 315M parameters). 
\vspace{-2mm}
\subsection{Whisper}
\vspace{-1mm}
Whisper \cite{radford2022whisper} is a family of sequence-to-sequence transformer \cite{vaswani17Transformer} models trained to perform multiple tasks such as multilingual ASR, language identification, and speech translation.
The models are pretrained on $\sim$680k hours of data retrieved from the world wide web and are available in five sizes between 39M parameters and 1.55B parameters. 
All models within the Whisper family share a similar encoder-decoder structure, differing only in parameters such as the number of transformer blocks and hidden layer dimensions.

The Whisper encoder $\mathcal{E}$ maps $J$ log-Mel spectrogram features obtained from the raw audio waveform $\boldsymbol{\hat{f}}(\boldsymbol{X})_{1:J} = \lbrace \boldsymbol{f}_1, \ldots , \boldsymbol{f}_J \rbrace$ to a sequence of $K$ hidden representations $\boldsymbol{H}_{1:K}$:
\vspace{-3mm}
\begin{equation*}
   \mathcal{E} : \boldsymbol{\hat{f}}(\boldsymbol{X})_{1:J} \mapsto \boldsymbol{H}_{1:K}. 
\vspace{-2mm}
\end{equation*}

The decoder autoregressively predicts the probabilities for the next token $y_i$, given the previous tokens $\boldsymbol{y}_{<i}$ and the hidden representations $\boldsymbol{H}_{1:K}$: $p(y_i \mid \boldsymbol{y}_{<i}, \boldsymbol{H}_{1:K})$. 
The model is trained on pairs of input spectrograms and target transcriptions using the cross-entropy objective. 
Due to the unavailability of ground truth transcriptions for all datasets in this study, adapting an ASR system to the domain is not straightforward. Therefore, we rely on Whisper for its strong multilingual performance on standard benchmarks, without the need for task-specific finetuning on downstream tasks \cite{radford2022whisper}. 
\vspace{-2mm}
\subsection{Low-rank Adaptation}
\vspace{-1mm}
In low-rank adaptation (LoRA) \cite{hu2022lora}, the pretrained weights of the underlying model are frozen and small trainable adapters are optimized instead. 
LoRA adapters utilize low-rank decomposition matrices $\boldsymbol{A} \in \mathbb{R}^{r \times n}$ and $\boldsymbol{B} \in \mathbb{R}^{m \times r}$ to define an incremental update $\boldsymbol{\Delta} \in \mathbb{R}^{m \times n}$, which are applied to the pretrained weight matrices of the underlying model. 
The updates to a weight matrix $\boldsymbol{W}$ are expressed as:
\vspace{-2mm}
\begin{equation*}
    \boldsymbol{W}^{'} = \boldsymbol{W} + \boldsymbol{\Delta} = \boldsymbol{W} + \boldsymbol{B} \boldsymbol{A}, 
\vspace{-2mm}
\end{equation*}
where the rank $r \ll \left\{n, m\right\}$, is a hyperparamter determining the dimensionality of the low-rank decomposition matrices, and thus, the quantity of trainable parameters within the module.
\vspace{-2mm}
\subsection{Minimum Bayes Risk Decoding}
\vspace{-1mm}
Minimum Bayes risk (MBR) decoding has long been used in ASR \cite{stolcke97_eurospeech,goel1998mbr,GOEL2000mbr} and machine translation (MT) \cite{kumar2002mbr_mt, kumar2004mbr_mt} as a means to improve accuracy of transcriptions or translations by considering not only the most likely hypothesis, but also the potential utility (or risk) associated with alternative hypotheses.

ASR systems typically utilize maximum a posteriori probability (MAP) decoding \cite{bahl1983map}, which maximizes the probability of selecting the correct word sequence \cite{stolcke97_eurospeech}. 
Since exhaustive MAP decoding is intractable, beam search is employed as an approximation instead \cite{aubert2002decoding}. 
However, MAP decoding exhibits several shortcomings that have been observed in both ASR and MT applications, such as a bias towards incomplete transcriptions \cite{chorowski2016better}, degrading performance with longer sequences \cite{cho-etal-2014-properties} and hallucinations due to low robustness under domain shift \cite{muller-etal-2020-domain}. 
We employ MBR decoding not only in the hope to mitigate some of the shortcomings of MAP decoding, but also to add more lexical diversity, by presenting a sequence of multiple hypothesis candidates to the LLM.

Let $\mathcal{U}(\boldsymbol{y^\prime}, \boldsymbol{y})$ be a utility function that compares an hypothesis string $\boldsymbol{y^\prime} \in \boldsymbol{\mathcal{Y}}$ against a reference string $\boldsymbol{y} \in \boldsymbol{\mathcal{Y}}$ from the space of all possible hypotheses $\boldsymbol{\mathcal{Y}}$. 
The optimal decision $\boldsymbol{y}_{opt} \in \boldsymbol{\mathcal{Y}}$ is the one that maximizes the expected utility (or minimizes the expected risk) for data generated under the model $p(\boldsymbol{y} \mid \boldsymbol{x}, \theta)$ \cite{eikema2020mbr}:
\vspace{-3mm}
\begin{equation*}
    \boldsymbol{y}_{opt} =\underset{\boldsymbol{y^\prime} \in \boldsymbol{\mathcal{Y}}(\boldsymbol{x})}{\operatorname{argmax}}\, \mathbb{E}_{p(\boldsymbol{y} \mid \boldsymbol{x}, \theta)}[\mathcal{U}(\boldsymbol{y^\prime}, \boldsymbol{y})].
    \vspace{-2mm}
\end{equation*}
Similar to MAP decoding, calculating the expectation over the entire space of all potential hypotheses $\boldsymbol{\mathcal{Y}}$ is often impractical, thus requiring the use of a tractable subset $\boldsymbol{\mathcal{\hat{Y}}} \subset \boldsymbol{\mathcal{Y}}$ of the full hypothesis space.
Previous works use n-best lists from beam search to compute a biased estimate of expected utility \cite{GOEL2000mbr,goel1998mbr,kumar2004mbr_mt,stahlberg2017mbr}. 
More recently, unbiased estimates have been obtained via Monte Carlo (MC) sampling \cite{eikema2020mbr,muller2021mbr,eikema2022sampling}. 

In MC sampling-based approaches, the set of possible hypotheses is approximated by drawing $S$ independent samples from the model via ancestral sampling.
For an hypothesis $\boldsymbol{y^\prime}$, the expectation is then maximized over $\boldsymbol{\hat{\mathcal{Y}}}$:
\vspace{-3mm}
\begin{equation*}    
    \boldsymbol{y}_{MC} = \underset{\boldsymbol{y^\prime} \in \boldsymbol{\hat{\mathcal{Y}}}(\boldsymbol{x})}{\operatorname{argmax}} \frac{1}{S} \sum_{s=1}^S \mathcal{U}(\boldsymbol{y^\prime}, \boldsymbol{y}_s) \text { with } \boldsymbol{y}_s \sim p(\boldsymbol{y} \mid \boldsymbol{x}, \theta). 
    \vspace{-2mm}
\end{equation*}
We employ the MBR method proposed in \cite{eikema2020mbr} with $S=10$ and the negated word error rate for $\mathcal{U}$ to conduct our experiments. 
 
\vspace{-2mm}
\subsection{Phonetic Transcriptions}
\vspace{-1mm}
Dysfluencies such as sound repetitions may not be visible in orthographic transcriptions and limit the benefit of lexical features based on word-level ASR systems. 
To analyze the effectiveness of more fine-grained units, we explore the phone recognition system described in \cite{xu2021w2v2phoneme} and predict phonetic transcriptions as an alternative to the orthographic transcriptions generated by Whisper.\footnote{We intentionally avoid the widely used but imprecise terms ``phonemic transcription'' and ``phoneme recognizer'' here.}
The phone recognition system is based on the 315M parameter version of wav2vec~2.0 pretrained on data from 53 languages \cite{conneau2020unsupervised} and finetuned on multilingual Common Voice data \cite{ardila2020commonvoice}.
It operates on symbols from the International Phonetic Alphabet as modeling units. 
The phonetic transcriptions are generated via greedy search and treated in the same way as the orthographic transcriptions, i.e., they are mapped to token identifiers and passed to the LLM (cf. Section~\ref{ssec:ours}). 
\vspace{-2mm}
\subsection{Our Approach}\label{ssec:ours}
\vspace{-1mm}
\begin{figure}[t]
  \centering
  \includegraphics[width=\linewidth]{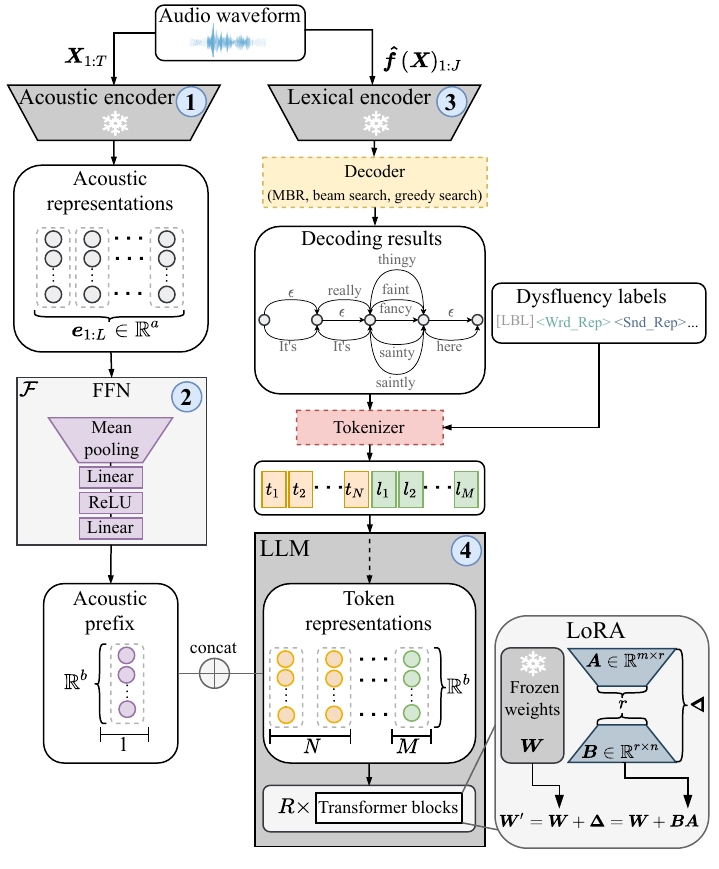}
  \vspace{-8mm}
  \caption{Overview of the components in our dysfluency detection system. The left segments illustrate the extraction and processing of acoustic features. The right segments demonstrate the retrieval and processing of lexical features, as well as the feature combination within the LLM backbone.}
  \label{fig:architecture}
  \vspace{-6.5mm}
\end{figure}
The overall composition of our system is inspired by \cite{deshmukh2023pengi,wagner2023multimodal,gong2024listen} and has four core components (cf. Figure \ref{fig:architecture}). 
The first component \circled{1} is an acoustic encoder, which generates latent representations from the input utterance. 
The model yields $L$ acoustic representations $\boldsymbol{e}_{1:L} \in \mathbb{R}^{L \times a}$ for each raw audio waveform input $\boldsymbol{X}_{1:T}$. 

The second main component  \circled{2} is a small feedforward network (FFN) $\mathcal{F}: \mathbb{R}^{L \times a} \mapsto \mathbb{R}^{b}$, consisting of mean pooling, two linear layers and ReLU activation. 
The FFN aggregates the extracted acoustic information along the time dimension and projects the result into the latent token space of a causal LLM.
The acoustic projection $\mathcal{F}(\boldsymbol{e}_{1:L})$ is concatenated with the token representations of the LLM. 
The combined representations are then passed through each of its $R$ Transformer blocks. 

The third main component \circled{3} is a pretrained ASR system, which is used to extract lexical features from the audio input by means of three different decoding methods (greedy search, beam search, and MBR decoding). 
The decoding algorithm generates viable hypothesis candidates, which are then tokenized and passed as inputs to the LLM. 
Beam search and MBR decoding yield multiple hypotheses, which are flattened to a string of words (i.e., all candidates are concatenated beginning with the most likely one). 

Acoustic and lexical features are combined in the fourth main component \circled{4}, a pretrained causal LLM. 
The LLM generates text tokens conditioned on the acoustic representations transformed via $\mathcal{F}$ and the ASR hypotheses candidates.
The weights of the LLM remain frozen during training and LoRA modules are optimized instead. 
During training, the weights of the LLM are held constant, while optimization focuses on the LoRA modules. 
Components \circled{2} and \circled{4} undergo joint optimization and components \circled{1} and \circled{3} remain unchanged throughout the training process.

For a training set containing $D$ acoustic representations, tokenized ASR hypotheses, and dysfluency label tokens $\{(\boldsymbol{e}_{1:L}^{(d)}, \boldsymbol{t}_{1:N}^{(d)}, \boldsymbol{l}_{1:M}^{(d)})\}_{d=1}^D$, the training objective for the parameters $\theta$ is the autoregressive prediction of the next dysfluency token $l_i$, given the previous dysfluency tokens $\boldsymbol{l}_{<i}$, the acoustic representations $\boldsymbol{e}_{1:L}$ and the ASR hypothesis tokens $\boldsymbol{t}_{1:N}$ using cross-entropy loss:
\vspace{-3mm}
\begin{equation*}
    \mathcal{L}_{\theta} =-\sum_{i=1}^M p_{\theta}\left(l_i \mid \boldsymbol{l}_{<i}, \boldsymbol{e}_{1:L}, \boldsymbol{t}_{1:N}\right).
    \vspace{-2mm}
\end{equation*}
At inference time, we use greedy search to generate a token sequence for a predefined maximum number of steps or until the \texttt{[EOS]} token is encountered. 
Dysfluency labels are then extracted from the generated text string. 
Any incomplete or inaccurately generated labels are discarded. 
We use a maximum number of 20 steps in our experiments. 

\vspace{-2mm}
\subsection{Modeling and Architecture Details}
\vspace{-1mm}
\begin{table*}[t]
  \caption{Multilabel dysfluency detection results using multilingual acoustic representations. Mod = Modified speech, Blk = Block, Int = Interjection, Pro = Prolongation, Snd = Sound repetition, Wrd = Word repetition. ``Finetuned'' indicates whether the acoustic features were domain-adapted or not. ``Layer'' refers to the wav2vec 2.0 transformer layer used for acoustic feature extraction. The column ``ASR Decoder'' shows the various types of decoding algorithms used to obtain ASR transcriptions.}
  \label{tab:exp}
  \vspace{-3mm}
  \setlength{\tabcolsep}{6.5pt}
    \scalebox{0.75}{
    \centering
 \begin{tabular}{c|c|c|c||c|c|c|c|c||c|c|c|c|c||c|c|c|c|c|c }
    \toprule
    \multirow{2}{*}{\textbf{Exp.}} & \multicolumn{2}{c|}{\textbf{Acoustic Features}} & \multirow{2}{*}{\makecell{\textbf{ASR} \\ \textbf{Decoder}}} & \multicolumn{5}{c||}{\textbf{\sepE}} & \multicolumn{5}{c||}{\textbf{\fb}} &  \multicolumn{6}{c}{\textbf{\ksof}} \\
    & Finetuned & Layer &   & \textbf{Blk} & \textbf{Int} & \textbf{Pro} & \textbf{Snd} & \textbf{Wrd} & \textbf{Blk} & \textbf{Int} & \textbf{Pro} & \textbf{Snd} & \textbf{Wrd} & \textbf{Mod} & \textbf{Blk} & \textbf{Int} & \textbf{Pro} & \textbf{Snd} & \textbf{Wrd} \\
    \midrule
    \hline
    \multicolumn{4}{c||}{Baseline \footnotesize{(experiments \#22-24 in \cite{bayerl23_interspeech})}} & \textit{0.32} & \textit{\textbf{0.77}} & \textit{0.53} & \textit{0.53} & \textit{\textbf{0.64}} & \textit{0.36} & \textit{0.79} & \textit{\textbf{0.62}} & \textit{0.64} & \textit{\textbf{0.52}} & \textit{0.75} &  \textit{\textbf{0.64}} & \textit{\textbf{0.85}} & \textit{\textbf{0.60}} & \textit{0.48}  & \textit{\textbf{0.14}} \\
    \hline
    \hline
    1               & \multirow{4}{*}{\xmark} & \multirow{4}{*}{12} & 1-best & \textbf{0.62} & 0.62 & 0.51 & 0.45 & 0.31 & 0.47 & 0.66 & 0.49 & 0.47 & 0.40 &  0.74 & 0.50 & 0.49 & 0.34 & 0.38 & 0.03 \\
    2               && & N-best & 0.61 & 0.60 & 0.51 & 0.44 & 0.31 & 0.40 & 0.65 & 0.46 & 0.47 & 0.36 &  0.76 & 0.51 & 0.45 & 0.27 & 0.34 & 0.04 \\
    3               && & Phon   & 0.60 & 0.58 & 0.50 & 0.42 & 0.30 & 0.53 & 0.69 & 0.41 & 0.50 & 0.42 &  0.77 & 0.47 & 0.49 & 0.21 & 0.29 & 0.04 \\
    4               && & MBR    & 0.61 & 0.62 & 0.52 & 0.40 & 0.31 & 0.47 & 0.63 & 0.46 & 0.47 & 0.37 &  0.78 & 0.48 & 0.50 & 0.18 & 0.26 & 0.06  \\
    \hline
    \hline
    5               & \multirow{4}{*}{\xmark} & \multirow{4}{*}{24}  & 1-best & 0.60 & 0.52 & 0.46 & 0.12 & 0.13 &  0.43 & 0.65 & 0.28 & 0.35 & 0.31 &  0.70 & 0.51 & 0.27 & 0.15 & 0.30 & 0.04 \\
    6               & & & N-best & 0.59 & 0.51 & 0.47 & 0.16 & 0.06 & 0.43 & 0.59 & 0.28 & 0.33 & 0.30  &  0.68 & 0.46 & 0.31 & 0.18 & 0.28 & 0.00 \\
    7               & & & Phon   & 0.59 & 0.52 & 0.43 & 0.21 & 0.10 & 0.49 & 0.61 & 0.34 & 0.39 & 0.15  &  0.70 & 0.46 & 0.28 & 0.16 & 0.30 & 0.01 \\
    8               & & & MBR    & 0.58 & 0.52 & 0.43 & 0.21 & 0.05 & 0.42 & 0.62 & 0.26 & 0.33 & 0.29  &  0.69 & 0.47 & 0.32 & 0.10 & 0.28 & 0.01 \\
    \hline
    \hline
    9               & \multirow{4}{*}{\cmark} & \multirow{4}{*}{12} & 1-best & 0.59 & 0.59 & 0.50 & 0.41 & 0.31 & 0.47 & 0.66 & 0.35 & 0.45 & 0.33 &  0.73 & 0.37 & 0.44 & 0.30 & 0.32 & 0.03 \\
    10              &  &  & N-best & 0.61 & 0.57 & 0.50 & 0.40 & 0.29 & 0.48 & 0.64 & 0.39 & 0.44 & 0.33 &  0.70 & 0.44 & 0.39 & 0.20 & 0.27 & 0.00 \\
    11              &  &  & Phon   & 0.60 & 0.57 & 0.51 & 0.40 & 0.27 & 0.52 & 0.64 & 0.38 & 0.46 & 0.39 &  0.73 & 0.45 & 0.44 & 0.29 & 0.31 & 0.06 \\
    12              &  &  & MBR    & 0.60 & 0.58 & 0.50 & 0.39 & 0.29 & 0.52 & 0.66 & 0.42 & 0.46 & 0.36 &  0.72 & 0.43 & 0.35 & 0.20 & 0.32 & 0.01 \\
    \hline
    \hline
    13               & \multirow{4}{*}{\cmark} & \multirow{4}{*}{24} & 1-best & 0.57 & 0.74 & \textbf{0.56} & \textbf{0.54} & \textbf{0.64} & 0.57 & \textbf{0.81} & 0.55 & \textbf{0.66} & 0.43 &  0.77 & 0.59 & \textbf{0.85} & 0.52 & 0.48 & 0.11 \\
    14               &  & & N-best & 0.57 & 0.73 & \textbf{0.56} & 0.53 & 0.61 & 0.56 & 0.80 & 0.56 & 0.61 & \textbf{0.52} &  0.77 & 0.55 & 0.84 & 0.49 & 0.43 & 0.04 \\
    15               &  & & Phon   & 0.56 & 0.72 & 0.54 & 0.48 & 0.58 & 0.56 & 0.79 & 0.52 & 0.63 & 0.45 &  \textbf{0.79 }& 0.56 & 0.82 & 0.48 & \textbf{0.50} & 0.08 \\
    16               &  & & MBR    & 0.58 & 0.73 & \textbf{0.56} & 0.49 & 0.56 & \textbf{0.58} & 0.78 & 0.48 & 0.63 & 0.44 &  0.78 & 0.57 & 0.82 & 0.52 & 0.49 & 0.12 \\
    \bottomrule
  \end{tabular}
  } 
  \vspace{-6mm}
\end{table*}
Each system is trained until convergence with an effective batch size of 32. 
Early stopping is applied, when the loss on the development set is not improving for five consecutive epochs. 
Each model used in Section~\ref{sec:exp} is selected based on the lowest loss achieved on the development set.
For optimization, we use AdamW \cite{loshchilov2018decoupled} ($\lambda = 10^{-4}$, $\epsilon = 10^{-8}$, $\beta_1 = 0.99$, $\beta_2 = 0.999$) with an initial learning rate of $2 \times 10^{-4}$, a linear schedule and a warm-up phase of 5\% of total training steps. 
The text token sequences are padded to a length of 1024. Token sequences longer than 1024 are truncated. 
Truncation occurred only in very few cases ($<1$\% of the training data). 
The hidden layers of $\mathcal{F}$ employ 512 units and are trained with a dropout probability of 10\%. 
We set $r = 64$ and the scaling factor for adjusting the magnitude of the adaption $\alpha = 16$ in all our experiments. 
The LoRA modules are optimized with a dropout probability of 10\%. 
We use the 7B parameter version of Llama 2 as the LLM backbone and a pretrained 315M parameter wav2vec~2.0 model as the acoustic encoder. 
The experiments on \sepE and \fb employ an acoustic encoder pretrained on the XLSR-53 \cite{conneau2020unsupervised} dataset, whereas the experiments on \ksof use a model pretrained on XLSR-53 that was subsequently finetuned on the German portion of Common Voice. 
The lexical encoder is either a pretrained 1.55B parameter Whisper \cite{radford2022whisper} model or the phone recognition system based on wav2vec 2.0 \cite{xu2021w2v2phoneme}. 
\vspace{-3mm}
\section{Experiments}\label{sec:exp}
\vspace{-2mm}
\subsection{Experimental Setup}
\vspace{-1mm}
We extracted acoustic representations from domain-adapted wav2vec 2.0 models using the method described in \cite{bayerl23_interspeech}, as well as their corresponding vanilla (i.e., out-of-domain) equivalents.
The acoustic representations were extracted at the \nth{12} and \nth{24} transformer layer of the wav2vec 2.0 system. 
Lexical features were either, phone-level (\textit{Phon}) or word-level 1-best hypotheses (\textit{1-best}) generated via greedy search, n-best lists generated via beam search (\textit{N-best}), or hypothesis candidates obtained via MBR decoding (\textit{MBR}). 
Beam search was configured with a beam width 12 and the number of hypotheses in each n-best list was 10. 
The baseline systems are the multilingual finetuned models from \cite{bayerl23_interspeech}. 
We report F1-scores balanced between precision and recall for all dysfluency types, as well as modified speech, which is only available for the \ksof corpus. 

Preliminary experiments involved utilizing either solely acoustic or solely lexical features for the dysfluency detection task. 
Lexical features alone underperformed compared to the use of acoustic-only features, and either standalone approach was outperformed by the combination of both acoustic and lexical features. 
Furthermore, we conducted experiments substituting Llama~2 with Falcon~7B \cite{almazrouei2023falcon} and Mistral~7B \cite{jiang2023mistral}, but no substantial performance differences were observed. 
We also examined different methods of aggregating acoustic feature sequences. 
However, using mean pooling to reduce the sequence length to one and passing a single aggregated acoustic vector as the prefix to the LLM performed comparably to using various aggregation levels that generate longer sequences. 
\vspace{-3mm}
\subsection{Results and Discussion}
\vspace{-2mm}
The results are summarized in Table~\ref{tab:exp}. 
Our LLM-based approach either matched or surpassed the baseline for most dysfluency types, when domain-adapted acoustic features extracted at the last layer of the finetuned wav2vec~2.0 system were used (cf. exp. \#13-16). 
The least overall improvement was observed on the German \ksof corpus, where our system using phonetic transcriptions only marginally surpassed the baseline for modified speech and sound repetitions (cf. exp. \#15). 
This discrepancy may be attributed to the predominance of English pretraining data for the LLM backbone, which may have more difficulty capturing the nuances of the German language. 

Our systems exhibit strong performance in identifying blocks within both the \sepE and \fb datasets with maximum relative improvements over the baseline of $\sim$94\% and $\sim$61\%, respectively. 
These improvements are consistent across all different types of acoustic features used. 
However, the improvement diminished on \ksof (at most $\sim$5\% relative), where the baseline performance for blocks is considerably stronger with an F1-score of 0.75. 

Generally, n-best lists and MBR hypotheses candidates did not yield significant enhancements over 1-best hypotheses. 
Only on the \fb dataset, n-best lists exhibited considerably better performance for word repetitions, surpassing phonetic transcriptions by $\sim$15\% relative (cf. exp. \#14). 
We initially anticipated that the LLM would benefit from more lexical alternatives and consider their variations. 
However, it appears that the relevant information extractable from a text sequence is already captured by 1-best hypotheses in most cases. 

Our approach did not lead to improvements on word repetitions. 
We hypothesize that Whisper's weakly supervised pretraining method may not have been the ideal option for modeling word repetitions in the lexical domain, as it may prioritize preserving the overall meaning of utterances, potentially leading to omissions of tokens with minimal additional information rather than generating exact word-for-word transcriptions.

Employing non-finetuned acoustic features from the last layer of wav2vec~2.0 yielded worse F1-scores compared to using the \nth{12} layer, regardless of the dataset (cf. exp. \#1-4 and \#5-8). 
This aligns with previous studies \cite{montacie2022stutter,bayerl22b_interspeech}, which indicate that transformer layers in the middle are more adept at capturing stuttering patterns. 
However, acoustic features from domain-adapted models yielded the best performance when extracted at the \nth{24} layer (cf. exp. \#13-16). 
As the model is finetuned, the layers closer to the output become more specialized to the specific task. 
Finetuning adjusts the parameters of these layers to better fit the characteristics of the dataset, yielding representations better suited to detecting stuttering patterns. 

\vspace{-3mm}
\section{Conclusions}
\vspace{-2mm}
Inspired by the recent advancements in LLMs, we approached multi-label dysfluency detection as a language modeling problem.  Our system jointly learns from a combination of acoustic and lexical features. 
Experimental results demonstrated that the Llama~2 backbone effectively combines acoustic and lexical information, matching or outperforming a robust baseline on the majority of dysfluency types across three datasets of English and German stuttered speech. 
We found that domain-adapted acoustic features from the last layer of a wav2vec~2.0 system yielded the best performance, particularly when combined with 1-best ASR hypotheses generated via greedy decoding. 
Future work will explore lexical encoder alternatives, the impact of varying LLM sizes and full end-to-end finetuning of both the LLM backbone and the acoustic encoder. 
\vspace{-1mm}
\section{Acknowledgments}
\vspace{-1mm}
We gratefully acknowledge the scientific support and HPC resources provided by the Erlangen National High Performance Computing Center (NHR@FAU) of the Friedrich-Alexander-Universität Erlangen-Nürnberg (FAU) under the NHR project b196ac14. NHR funding is provided by federal and Bavarian state authorities. NHR@FAU hardware is partially funded by the German Research Foundation (DFG) – 440719683.
This work was supported by the Bavarian State Ministry of Science and the Arts under grant H.2-F1116.NÜ/61/2.
\FloatBarrier
\bibliographystyle{IEEEtran}
{\footnotesize{
    \bibliography{refs}
}

\end{document}